# Transfer learning for financial data predictions: a systematic review

Vincenzo Lanzetta,[1]

1 Information Technology and Electrical Engineering Department, University of Naples Federico II

To whom correspondence should be addressed; E-mails: vincenzo.lanzetta@unina.it

**Abstract**

Literature highlighted that financial time series data pose significant challenges for accurate stock price prediction, because these data are characterized by noise and susceptibility to news; traditional statistical methodologies made assumptions, such as linearity and normality, which are not suitable for the non-linear nature of financial time series; on the other hand, machine learning methodologies are able to capture non linear relationship in the data. To date, neural network is considered the main machine learning tool for the financial prices prediction. Transfer Learning, as a method aimed at transferring knowledge from source tasks to target tasks, can represent a very useful methodological tool for getting better financial prediction capability. Current reviews on the above body of knowledge are mainly focused on neural network architectures, for financial prediction, with very little emphasis on the transfer learning methodology; thus, this paper is aimed at going deeper on this topic by developing a systematic review with respect to application of Transfer Learning for financial market predictions and to challenges/potential future directions of the transfer learning methodologies for stock market predictions.

# 1. Introduction

The theory of Efficient Market Hypothesis (EMH) [40] has been developed with the aim to understand how financial market prices change over time; according to EMH, financial markets immediately adjust their prices to the information available [1]. In an efficient market a security's price fully reflects all available information about the intrinsic security value [2]. In such a market the securities prices are always fair (i.e. correctly incorporate all information available up to the current time) and there aren't mispriced securities; as consequence, in an efficient market it is not possible to exploit mispriced securities in order to gain extra returns with respect to the fair returns.

EMH would requires the presence of only "active" investors which are able to immediately incorporate information into prices; but in an efficient market nobody could gain greater profits - than the index market profits - because of the exclusive presence of fair prices corresponding to the intrinsic securities values; in other words, in an efficient market there aren't extra profits to be gained because of the absence of mispriced securities; as consequence, in an efficient market there is no benefit to be "active" investor and the smart investors should only be "passive" ones that only follow the "buy and hold" strategy; but the presence of only "passive" investors lead to a paradox because nobody would immediately incorporate information into prices and - consequently - financial markets would not be able to immediately adjust their prices to the information available; so, the above paradox implies market cannot be efficient, at least in order to let the existence of smart "active" investors which should be able to incorporate information into prices; moreover, the above necessary existence of smart "active" investors also implies the presence of mispriced securities and – as consequence – the existence of the prices predictability by means of the gap exploiting between the mispriced securities and the fair prices.

Because of the predictability of financial prices, the market prices have to follow a non-random walk; obviously, the more efficient is the market, the less predictable are the price changes and – as consequence – the more random are the sequence of price changes generated by such a market [3].

According to Mishkin & Eakins [2], lots of reasons are able to explain the inefficiency of financial markets: 1) information are not available to all the people in the same moment, 2) different financial operators could be in disagree with the information effect on prices and on expected returns; 3) transaction costs and taxes impact in a different manner on different operators; 4) overconfidence and social contagion are able to push the prices far from the actual intrinsic value. Accordingly, to the above reasons, literature highlights empirical evidences contrary to the efficiency of financial markets [1]: volatility surging or bubbles prices are empirical notorious examples [2].

With the aim to overcome the limits of EMH, the Adaptative Market Hypothesis (AMH) states that <<prices reflect as much information as dictated by the combination of environmental conditions and the number and nature of "species" in the economy>> [3]; as consequence of the prices dependence on dynamics between financial operators and market conditions, smart investors are able to exploit the "temporary" market inefficiencies and - thus - are able to gain extra profits with respect to the actual intrinsic securities values; but, notwithstanding the financial market predictability due to the recognized financial market inefficiency, the prices are difficult to be predicted because of many interrelated variables - such as economic, political, psychological, and company-specific ones [4] – and because of their non-stationary nature [1, 68]; more on this point, Nguyen et al. [68] have highlighted that financial time series data pose significant challenges, for accurate stock price prediction, because these data are characterized by noise and susceptibility to news; furthermore, the above researchers also highlighted the non suitability of traditional statistical methods - like weighted moving average (WMA), generalized autoregressive conditional heteroskedasticity (GARCH), and autoregressive integrated moving average (ARIMA) - because the traditional methodologies made assumptions, such as linearity and normality, which are not suitable for the non-linear nature of financial time series.

Recent literature, by highlighting new and successful approaches - for financial predictions - characterized by the adoption of the so called <<technical analysis>> within econometrics, statistics, data mining, and artificial intelligence methodologies [4] - states that machine learning (ML) is the << logical next step of the finance researcher toolkit>> [9], as ML methodologies are able to capture non linear relationship in the data [7].

ML literature for financial market prediction is vast [1], increasing ([9] and with a wide range of topics/categories [9] and techniques [7] being successfully applied for investment strategies and prices/returns prediction; this very large body of literature have used – as data source - several security's data in order to get results; among them, forex (acronym for foreign exchange) data has been extensively used in ML papers [9]; indeed, as the critical determinants of the forex price movements are macro economical ones (for example, expected inflation, geopolitical tensions, commercial relationships between countries, public debts, deficits, interest rates), adopting the forex data - as data source - let to avoid the typical issues linked to the stock prices predictions, such as the false firms communications and the insider trading issues.

The forex market is the most liquid financial market, i.e. the biggest one from the point of view of the money's volume; as <<liquidity stimulates arbitrage activity, which, in turn, enhances market efficiency>>[17], it has to be noted that the great liquidity of forex markets could impact on predictability results because a greater liquidity push the market closer to the random walk behaviour, i.e. it push the market closer to the unpredictability; among all the forex currency pairs, the most traded one is EUR/USD [28].

With the aim to forecast exchange rates of currency pairs in the FOREX market, some researchers – by comparing the ARIMA popular econometrical model with several neural network (NN) architectures – stated that <<all NN models outperformed the ARIMA model>> [8]. To date, neural network is considered the main ML tool for the financial prices prediction [1, 8], mainly because of its ability to deal with complex and non-linear relationship of the data [7]; more in depth, a recent literature review [20] highlighted that <<deep learning is proven as the state-of-the-art technique for predicting the stock market in most of the surveyed studies>>; furthermore, recent literature stated that convolutional/recurrent based NN architectures have been extensively applied in order to get good financial market predictions from input such as fundamental accounting data or as technical indicators. [1,7].

Recent forecasting markets approaches combine the predictions of different ML models – acting each one as an individual trader - in a classification ensemble [8]; indeed, as <<financial data is notorious for having a low signal-to-noise ratio>> [7], combining methods represents a possible solution to the results instability: << to increase the stability of our NN model, for each training data set, 1000 NN models will be trained and the top 50% best performers will be pooled for prediction. We call this process as the ensemble NN>> [18]; accordingly, to the above approach, some authors have merged the predicted outputs - of several test set results - in order to get the overall NN performance [12].

Within the machine learning framework, transfer learning models are methods aimed at transferring knowledge from source tasks to target tasks, with the general aim to leverage the knowledge - of previously trained models - in order to achieve better performances than traditional ML models.

According to a growing literature interest on the transfer learning models for financial data predictions, I highlight the need to analyze the current state of the art because - to best of my knowledge - the existing reviews on the above body of knowledge is mainly focused on neural network architectures for financial prediction, with very little emphasis on the transfer learning methodology [63, 75]; furthermore, the current literature reviews on transfer learning for time series [69, 70] don't have a deep and broad focus on financial predictions.

Therefore, according to the growing body of recent studies that specifically explore different applications of transfer learning in the context of financial predictions, there is a distinct need to compile and organize the literature findings. This would involve creating a comprehensive review paper that addresses the relationship between transfer learning and financial forecasting.

Thus, I follow a systematic approach to review previous studies in order to understand how Transfer Learning has been applied for financial market predictions; accordingly, section 2 presents the relevant background aspects related to the so-called fundamental/technical analysis and with respect to the relative use within the machine learning field; the research questions are presented in section 3; the description of the followed systematic protocol is reported in section 4; section 5 presents the results of the conducted review; finally, conclusions are in section 6

## 2. Background

Two approaches [4] have been traditionally used, by the financial traders, for prices prediction: fundamental analysis and technical analysis. The fundamental analysis (FA) <<seeks economic factors that have an influence on market trends>> [1] and - in order to get predictions - takes in consideration macroeconomic analysis, industry analysis and company ones[ 4]; so, following a thorough examination of the fundamental factors, an investor has the opportunity to purchase a stock if the analysis reveals that the stock's fair value exceeds its current market price, indicating an undervalued condition. In such a situation, the investor acquires the stock and retains it until the market price aligns with its intrinsic value. Consequently, the investor realizes a substantial profit by selling the stock at its intrinsic market price; likewise, an investor might choose to sell - or abstain from purchasing - an overpriced security.

Fundamental analysis is one of the most effective ways to evaluate investments, as it involves examining every aspect of a company's operations through its balance sheet, past performance, financial reports, market goodwill, management, and consumer behaviour, with the final aim to define the intrinsic value of the related stock. The analysis usually begins from macroeconomic factors, such as the economy and industry performance, and goes down to microeconomic factors like management, strategic initiatives, and business policies. Thus, FA may guide investors to manage risks and make informed investing decisions by ascertaining the intrinsic value of a stock.

The technical analysis - highly popular for its potential usefulness with respect to the forecasting of the stock market risk premium [11] - considers <<past and current market price, trading volume and other publicly available information to predict future market prices.>> [5], and revolves around the meticulous observation of stock price fluctuations in order to forecast future trends and facilitate investment decision-making.

As the economic fundamentals are not practically able to sensibly impact on the shorter market dynamics, the shorter is the horizon, the more used is the technical analysis: <<at short horizons (less than a week) traders use technical analysis more frequently than fundamental analysis.>> [6]; on this last regard, a financial trader needs to define a trading strategy - which is constituted by rules for entering and exiting the trades, for risk management and for money management [19] - by selecting a specific operative trading timeframe depending on a trade-off, within a trading session, between the potential net profit (including transaction costs and taxes) of each trade and the potential maximum numbers of trades; with respect to a trading session, shorter timeframes (such as 30 minutes, for example) - which are able to limit the risks due to the high volatility linked to macro economical communications or to rare and unpredictable events - could lead to get lots of trading signals, but could also lead to manage price movements not enough large to overcome the very frequent trading transaction costs; on the contrary, greater timeframes (1 day, for example) offer greater price movements, but limit the maximum trade numbers within a trading session.

**2.1 Fundamental analysis typologies**

Fundamental analysis encompasses two distinct approaches: qualitative analysis and quantitative analysis. Qualitative analysis focuses on intangible factors such as market conditions, brand reputation and overall goodwill. On the other hand, quantitative analysis is predominantly driven by statistical data and numerical metrics

Qualitative analysis entails examining various aspects of a company's operations, including its market reputation, consumer behaviour, demand dynamics, and overall recognition in the broader market. The objective is to uncover insights into how the company is perceived, how its management decisions or announcements generate market excitement, and what sets it apart from its competitors. Additionally, factors such as brand value and other relevant indicators shed light on the company's socio-economic standing within the market.

Quantitative analysis primarily focuses on utilizing statistical data, reports, and financial information. It relies solely on the examination of a company's financial statements, quarterly performance reports, balance sheets, debt levels, cash flow statements, and other numerical data. This approach involves analyzing various numbers, ratios, and values to gain insights into the stock's price and assess the overall financial well-being of the company.

**2.2 Fundamental analysis and machine learning**

In a recent review of literature [47] on machine learning studies applied to stock market prediction using fundamental, technical, and combined analyses, 66% of the reviewed documents are predominantly relied on technical analysis. In contrast, 23% and 11% of the studies are focused on fundamental analysis and combined analyses, respectively; furthermore, the findings revealed that out of the 122 papers reviewed, 28 of them utilized fundamental analysis for predicting stock market behaviour. Interestingly, 98% of the last studies employed sentiment analysis, of social network sites, as predictors of market movements; on the other hand, the review highlighted a significant gap in research regarding the usage of macroeconomic variables, as the sole data source for stock market prediction, despite some authors recognizing a positive correlation between macroeconomic variables and stock market returns; Therefore, as this area remains largely unexplored in the existing literature, authors of the above review concluded that there is a clear need for future research to investigate the prediction of stock market trends based on macroeconomic variables[.]

**2.3 Technical analysis typologies**

Technical analysis is a set of financial prediction techniques derived by the analysis of time series history of the particular asset price, either graphically or mathematically [15].

Although technical analysis is not based on an underlying economic or financial theory, it is widespread used among traditional financial practitioners in financial markets - in general - and in the forex market - in particular - mainly in combination with fundamental analysis [15, 16].

Technical indicators calculated from historical time series data, or line charts based on security prices, or candlestick chart developed from historical data, are the main financial prediction techniques of the technical analysis.

Technical indicators, which are statistical tools aimed at identifying buy or sell signals, are calculated from historical time series data with the use of specific formulas aimed at incorporating the historical price behaviour in a single value; the above formulas are based on prices and money volumes related to past trading periods, usually no more than 20–30 trading periods [8]; because of the very large number of possible parameter combinations within the technical indicators formulas, it is possible to obtain a very large number of possible trading rules; as example, in a large-scale investigation of technical trading rules of the foreign exchange market, Hsu et al. [15] examining over 21,195 distinct technical trading rules, constituted by <<2,835 filter rules, 12,870 moving average rules, 1,890 support-resistance signals, 3,000 channel breakout rules and 600 oscillator trading rules>> [15]; so, because of the above very large number, it is not easy to select the financial trading rules to be used; moreover, the more technical trading rule are selected, the greater is the relative correlation risk and the less informative could became the relative trading rules portfolio; so, traditional financial practitioners - which basically use each single trading rule, of their trading rule portfolio, as a single signal rule - are subject to lots of difficulties for building their trading rules portfolio, and generally limit the number of selected trading rules as answer aimed at managing the above correlation risk.

The drawing of line charts, i.e. the drawing of financial prices line, is a method used by traditional financial practitioners for discovering the points where prices tend to react. For traditional financial practitioners the chart patterns – which are considered as the footprints of the smart money – are considered as an useful tool for predicting where the stock price is heading, how far it will travel, and how reliable is the trail [38]. To this aim the financial practitioners - which try to find particular price patterns, within the analysed price line charts, that usually lead to a predictable price movement - have assigned to each particular pattern a name related to the relative shape, such as "pennants", "flags", "head and Shoulder", "triangles" and so on [37].

A candlestick chart – which is a combination between a line chart and a bar chart – is used by traditional financial practitioners to describe the price movements for a given period of time and with respect to a given time-step. In a candlestick chart there are a selected number of candles, related to a specific time-step (30 minutes, 1 hour, 1 day, 1 week and so on) and constituted by an upper line, a lower line and a real body; the number of candles, of a candlestick chart, defines the given period of time of the chart; the candle body height represents the price distance between the opening and the closing of the relative candle time-step. The real body is usually filled in red colour if the open price is higher than the closing price, or in green colour in the opposite case; the upper and lower lines, at the end of the real body, represents the high and low price range of the relative time-step, respectively [25].

## 2.4 Technical analysis and machine learning

Technical indicators are broadly used in financial machine learning literature because of the practical success they have demonstrated [9]; related literature highlights that the majority of existing ML-based financial trading methods uses technical indicators, instead of raw prices, as input data [8]; indeed, <<Using technical indicators can be more informative than using pure prices>>[10]; on this point, Nguyen et al. [68] highlighted that, while technical indicators have been widely used as input variables for stock market prediction, selecting suitable indicators remains a challenge.

With the aim to exploit the power of deep convolutional neural networks (CNNs) for an algorithmic trading system, some authors have converted 1D values of financial technical indicators into a 2D image-like data representation [12]; with the aim to directly use financial images which are commonly analysed by financial practitioners, scholars recently proposed some CNN-based architectures on price patterns of financial chart lines images [28,39] and on candlestick chart images [25, 26, 27, 28, 29, 30, 31, 32]; table 1 presents recent papers - on CNN-based architectures for financial candlestick images - in order to highlight literature interest on the above topic.

| f.# | Authors, year | Problem | NN architecture |
|---|---|---|---|
| [25] | Ho, T.-T.; Huang, Y., 2021 | Classification | 1D CNN joined with 2D CNN |
| [27] | Rosdyana Mangir Irawan Kusuma et al, 2019. | Classification | CNN |
| [28] | Moghaddam AH, Momtazi S, 2021 | Classification | CNN + CNN-LSTM |
| [29] | T. Kim, H.Y. Kim, 2019 | Regression | <<feature fusion LSTM-CNN>> |
| [30] | S. Birogul et al., 2020 | Classification | CNN (real-time object detection system -YOLO) |
| [31] | Hung, C.-C.; Chen, Y.-J., 2021 | Classification | CNN-autoencoder (CAE) - RNN with GRU gating mechanism |
| [32] | Zeng Z et al. 2021 | Classification | mirrored VGG16-LSTM |

Table 1 – recent literature on CNN-based architectures for financial candlestick images

## 2.5 overview on the current studies about transfer learning and financial predictions

Transfer Learning, as <<the ability of a system to recognize and apply knowledge and skills learned in previous tasks to novel tasks>>, is aimed at extracting << the knowledge from one or more source tasks and applies the knowledge to a target task>> [45]. Transfer learning can be categorized into three primary types: transductive transfer learning, inductive transfer learning, and unsupervised transfer learning [79]. Transductive transfer learning is exemplified by scenarios where labelled data is exclusively available in the source domain; in such cases, the source and target domains differ but are able to provide similar outer knowledge to support the learning process. In the context of transductive transfer learning, a pre-trained model gains insights from both the source domain and the source task; this acquired knowledge is subsequently leveraged to enhance the learning process of the target predictive function. In the context of inductive learning, knowledge transfer can be transferred within the same domain. The source domain's data may be either labelled or unlabelled, while data in the target domain are typically labelled to facilitate the creation of an objective predictive model. Furthermore, inductive transfer learning can be subdivided based on the data conditions in the source domain, resulting in two distinct cases: multi-task learning and self-taught learning. Multi-task learning comes into play when there is an abundance of labelled data in the source domain, coupled with a limited amount of labelled data in the target domain. On the other hand, self-taught learning is employed when only unlabelled data is accessible in the source domain. In this scenario, the pre-trained model utilizes unsupervised learning to acquire knowledge, and get fine-tuned with labelled data in the target domain. Unsupervised transfer learning comes into play when there is a lack of labelled data in both the source and target domains during training. The process involves utilizing unsupervised learning frameworks for both pre-training and retraining stages. In this approach, the chosen source and target tasks are distinct but related.

Notwithstanding the recent increasing number of studies, with good results, related to the transfer learning application within the field of the financial predictions, to the best of my knowledge the literature presents few reviews on neural network architectures, for financial prediction, with very little emphasis on the transfer learning methodology [63, 75], and few reviews on transfer learning for time series [69, 70] without a deep and broad focus on financial predictions.

## 3 Research questions

According to the recent literature interest on the transfer learning methodology within the field of the financial data prediction, and according to the current lack of specific literature reviews which are deeply and broadly focused on the above topic, this paper is aimed at answering to the following research questions:

1. How Transfer Learning has been applied for financial market predictions?
2. Which are the challenges and potential future directions of the transfer learning methodologies for financial market predictions?

## 4 Description of the systematic protocol

This section presents the adopted systematic protocol [44], according to specific and latest literature findings applied to the engineering field (Torres-Carrion et al., 2018, in [44]), aimed at selecting the relevant papers on the transfer learning methodologies for financial data predictions.

## 4.1 Preparing the systematic research:

Table 2 presents all the steps aimed at selecting the relevant papers, with the relative adopted criteria.

| Step number | Step description | adopted criteria |
|---|---|---|
| 1 | Defining the initial filter | article title, abstract, keywords |
| 2 | Defining the years | from current year-5 (i.e., 2019) to date |
| 3 | Defining the subject area | Computer Science, Engineering |
| 4 | Defining the search words and the relative synonymous | details are presented in Table 4 and Table 5 |
| 5 | Defining the exclusion criteria | no books, no technical reports, no paper published before current year-5 (i.e., 2019) |
| 6 | Defining the Journal selection criteria | journals with the highest ScimagoJR impact factor with respect to the following research area: machine learning, computer science, artificial intelligence; accordingly, the "aims and scope" declarations of each journal are considered as discriminating elements |
| 7 | Defining the columns headers of the data extraction form to be used for the organization of the results of the systematic search process | problem taxonomy, used dataset and related characteristics, methodology description, model architecture, main results, journal name, ScimagoJR journal classification, title, authors, year |

Table 2 – systematic search process with adopted criteria

## 4.2 Conducting the systematic research:

The systematic research has been conducted on the Scopus Database; by applying the process of section 4.1, and by means of an iterative and incremental process with respect to the search words (and the relative synonymous) of table 3 and table 4, I get the list of primary studies to be analysed.

In the following tables are presented all the details on the conducted systematic process; the asterisk (*) at the end of some words - of the following tables - has been used in order to generalize any possible symbol after the asterisk, according to the aim to increase the efficiency of the whole search process.

Thus, table 3 presents the initial search syntax aimed at looking for the eventual presence of reviews in the field of the transfer learning for financial predictions.

| Scopus | Initial filter: <Article Title, Abstract, Keywords> |
|---|---|
| | Year: >2018 |
| | Subject area: Computer Science, Engineering |
| | Search words and relative synonymous: review AND "transfer learning" AND (finance OR trading OR stock) |
| | Exclusion criteria: no books, no technical reports |
| | Journal selection criteria: journals with the highest ScimagoJR impact factor with respect to the following research area: machine learning, computer science, artificial intelligence; accordingly, the "aims and scope" declarations of each journal are considered as discriminating elements |

Table 3 – initial search syntax aimed at looking for the eventual presence of reviews in the field of the transfer learning for financial predictions

The output of the table 3 search criteria (extracted from SCOPUS on February 27, 2023 and on January 3, 2024) – combined with the one of the table 4 and with the one derived by an explorative non systematic search – is constituted by four reviews [63, 69, 70, 75] that are all characterized by the absence of a deep and broad focus on the topic of the transfer learning for financial predictions.

Table 4 presents the search syntax – extracted from SCOPUS on February 28, 2023 and on January 3, 2024 - for specific studies in the field of the transfer learning for financial predictions, and is aimed at developing the list of primary studies (table 5) to be analysed for their eventual direct evidence with respect to research questions.

| Scopus | Initial filter: <Article Title, Abstract, Keywords> |
|---|---|
| | Year: >2018 |
| | Subject area: Computer Science, Engineering |
| | Search words and relative synonymous: "transfer learning" AND (finance OR trading OR stock) |
| | Exclusion criteria: no books, no technical reports |
| | Journal selection criteria: journals with the highest ScimagoJR impact factor with respect to the following research area: machine learning, computer science, artificial intelligence; accordingly, the "aims and scope" declarations of each journal are considered as discriminating elements |

Table 4 – search syntax for specific studies in the field of transfer learning for financial predictions

figure 1 highlights the growing body of literature on all subject area, from 2010 to 2023, for "transfer learning" AND (finance OR trading OR stock) as search words in Scopus.com

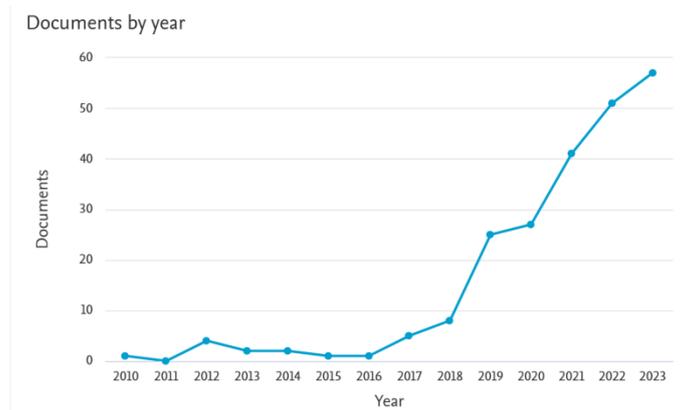

Figure 1: documents by year (all subject area, from 2010 to 2023, for "transfer learning" AND (finance OR trading OR stock) as search words in Scopus.com

## 5 Results of the conducted review
### 5.1 Summary of the reviewed papers

This section is aimed at providing the summary of the reviewed articles with respect to the following aspects: journal classification, problem taxonomy (regression, classification, rewarding, clustering, review), used dataset with related characteristics, model architecture, methodology description, main results, more in depth, in table 5 is presented - for each selected paper - journal classification, problem taxonomy, model architecture and transfer learning typology.

According to this study classification of the transfer learning typology for every analyzed study, Pan and Yang [82] defined 2 transfer learning typologies: i) Inductive transfer learning (ITL), in which source and target domains are the same (i.e.: source and target domains have the same features) and source and target task are different but related (i.e.: the label spaces of domains are different, or the conditional probability distributions of domains are different); ii) transductive transfer learning (TTL), in which source and target domains are different but related (i.e.: source and target domains have different but related features) and source and target task are the same (i.e.: the label spaces of domains are the same, or the conditional probability distributions of domains are the same). According to Pan and Yang [82], we also highlight that when target and source domains are the same and their learning tasks are the same, i.e., the learning problem has to be classified as a traditional machine learning problem.

| Article number | Authors | Journal classification | Problem taxonomy | Model architecture | Transfer Learning typology |
|---|---|---|---|---|---|
| 5-1 | Zhang D., Lin R., Wei T., Ling L., Huang J., 2023 | Q1 – Artificial Intelligence | Regression | Adversarial domain adaptation architecture | TTL |
| 5-2 | Sokolov A., Kim J., Parker B., Fattori B., Seco L., 2023 | Q2 – Artificial Intelligence | Regression | Representations of Interrelated Financial Time Series architecture | Not understandable (no features specification) |
| 5-3 | RC Chen, WI Yang, KC Chiu, 2022 | Conference paper | Regression | Long Short Term Memory Network (LSTM) | TTL |
| 5-4 | Pal S.S., Kar S., 2022 | Q2 geometry and topology | Regression | Fuzzy Transfer Learning (no NN architecture) | Fuzzy Transfer Learning |
| 5-5 | Otović E., Njirjak M., Jozinović D., | Q1 Artificial intelligenc | Regression and classification | CNN-LSTM | TTL |
| | Mauša G., Michelini A, Štajduhar I., 2022 | e | | | |
| 5-6 | Junwen Yang, Yunmin Wang, Xiang Li, 2022 | Q2 computer science | Classification | LASSO-LSTM | TTL |
| 5-7 | He, QQ., Siu, S.W.I. & Si, YW., 2022 | Q2 Artificial intelligence | Classification | LSTM with attention mechanism | TTL |
| 5-8 | Borrageiro G., Firoozye N., Barucca P. The Recurrent Reinforcement Learning Crypto Agent. 2022 | Q1 Computer Science (miscellaneous) | Rewarding | Recurrent Reinforcement Learning | ITL |
| 5-9 | Borrageiro G., Firoozye N., Barucca P. Reinforcement Learning for Systematic FX Trading. 2022 | Q1 Computer Science (miscellaneous) | Rewarding | Recurrent Reinforcement Learning | ITL |
| 5-10 | Yang Li, Hong-Ning Dai & Zibin Zheng, 2022 | Q2 Artificial intelligence | Classification | Attention-based LSTM network with adversarial training | ITL |
| 5-11 | Jiao Chen et al., 2021 | Q1 Artificial intelligence | Rewarding | Deep reinforcement learning algorithms | ITL |
| 5-12 | Badr Hirchoua et al., 2021 | Conference paper | Rewarding | Deep reinforcement learning | ITL |
| 5-13 | Haoran Xu et al., 2021 | Conference paper | Regression | LSTM | ITL |
| 5-14 | Chenkang Lv; Boyong Gao; Cui Yu, 2021 | Conference paper | Regression | Bidirectional LSTM | ITL |
| 5-15 | Shanshan Dong and Chang Liu, 2021 | Not yet assigned quartile | Classification | CNN | ITL |
| 5-16 | E. Fons, P. Dawson, X. -j. Zeng, J. Keane and A. Iosifidis, 2021 | Not yet assigned quartile | Classification | LSTM | ITL |
| 5-17 | Hai V. Che, Trung Q. D. Tran & Duc M. Duong, 2021 | Conference paper | Classification | LSTM | TTL |
| 5-18 | Qi-Qiao He, Patrick Cheong-Iao Pang, Yain-Whar Si, 2020 | Conference paper | Regression | LSTM | ITL |
| 5-19 | Adriano Koshiyama, Nick Firoozye, Philip Treleaven, 2020 | Conference paper | Review | Several reviewed architectures | - (review paper) |
| 5-20 | Kei Nakagawa, Masaya Abe, Junpei Komiyama, 2020 | Conference paper | Regression | Feed forward Neural Network | ITL |
| 5-21 | Jie Fang, Jianwu Lin, 2020 | Not yet assigned quartile | Classification | Feed forward Neural Network | TTL |
| 5-22 | R. Miao, X. Zhang, H. Yan and C. Chen, 2019 | Not yet assigned quartile | Classification | Bi-directional LSTM | TTL |
| 5-23 | Vasant Dhar et al., 2019 | Q2 Computer Science Applications | Classification | CNN | TTL |
| 5-24 | Thi-Thu Nguyen and Seokhoon Yoon, 2019 | Q2 Engineering (miscellan | Classification | LSTM | TTL |

| | | | | eous) | | |
|---|---|---|---|---|---|---|
| 5-25 | S. Merello, A. P. Ratto, L. Oneto and E. Cambria, 2019 | Not yet assigned quartile | Regression | Feed forward neural network | TTL |
| 5-26 | X. Man, T. Luo and J. Lin, 2019 | Not yet assigned quartile | Review | Several reviewed architectures | - (review paper) |
| 5-27 | Gyeeun Jeong et al., 2019 | Q1 Artificial intelligence | Rewarding | Reinforcement learning | ITL |
| 5-28 | Svetla Boytcheva and Andrey Tagarev, 2019 | Q2 Information Systems and Management | Clustering | Density based clustering method | - (clustering approach based on transfer learning idea) |
| 5-29 | Vibha Tripathi, 2019 | Not yet assigned quartile | Review | No compared architectures | - (review paper) |
| 5-30 | Nihal V. Nayak, Pratheek Mahishi, Sagar M. Rao, 2019 | Not yet assigned quartile | Classification | Bi-LSTM | TTL |

Table 5 – features of reviewed articles

In 5-1 [80] Zhang D. et al. have highlighted that recent advancements in deep learning models have demonstrated success in enhancing the accuracy of financial time-series forecasting, but - due to the implicit complexity of financial information and the scarcity of available labelled data - existing benchmarks in this domain exhibit poor generalization capabilities. To address the challenges posed by overfitting, authors introduced a novel deep transfer learning framework that incorporates adversarial domain adaptation specifically designed for financial time-series forecasting tasks. The approach involved the development of a standard adversarial domain adaptation architecture to transfer feature knowledge and mitigate the distribution discrepancies between different financial datasets. Zhang D. et al. pointed out the need to have a careful selection of the source domain as it significantly influences forecasting performance; accordingly, authors employed a <<temporal causal discovery method>> to appropriately select the source dataset. The feasibility and effectiveness of the proposed framework has been validated through empirical experiments, ablation studies,

Diebold–Mariano tests, and parameter sensitivity analyses conducted on various financial datasets from three domains, namely financial indexes, energy futures, and agricultural futures.

In 5-2 [81] Sokolov A. et al. have introduced a financial time-series representation model that combines a pretraining task and a neural network architecture to create generalized representations of multiple financial time-series inputs, with the final aim to predict future pairwise correlations among securities. To foster the learning of correlation structures among securities, the authors employ a Siamese network architecture, which is an identical network - characterized by a shared architecture and weights -, that is employed to encode two sets of input data in a pairwise manner.

In 5-3 [43] the authors have proposed a financial price prediction approach, by using Stock daily prices and macro-economical features, that leverages a transfer learning-based LSTM in order to address the challenge of the limited data availability. The experimental findings demonstrated that the transfer learning approach yields the most favourable outcomes, exhibiting the lowest error rates. This indicates that the use of transfer learning, in conjunction with LSTM neural networks, effectively improves the accuracy of financial price predictions.

In 5-4 [48] Pal et al. have utilized fuzzy transfer learning (FuzzyTL) logic for time series forecasting. To establish the framework of FuzzyTL, the authors first transformed the labelled dataset into a fuzzy representation. Then, they employed a de-fuzzification method to generate output predictions for the unlabeled dataset. To evaluate the effectiveness of the method, Pal et al. conducted a comparative study using three distinct daily stock datasets with high and low prices as input data. Findings indicated that combining knowledge transfer with the smoothing of dependent attributes produces favourable results, especially when dealing with fluctuating stock market data over time. In conclusion, this study suggests that the incorporation of knowledge transfer techniques can enhance the accuracy of stock market predictions.

In 5-5 [22] Otović et al. have emphasized that one potential solution to address the scarcity of data is employing transfer learning; so, they examined the conditions under which transfer learning proved effective - by using daily time series data with closing price in 50 consecutive workdays as features - within a transfer learning-based CNN-LSTM architecture. The authors repeated the experiments several times, and applied statistical tests with the aim to confirm the results validity. The Wilcoxon signed-ranks test, as statistical test, were used to address the multiple comparisons problem and to mitigate the risk of obtaining false discoveries; accordingly, to best practices related to the above test, Otović et al. employed a well-accepted two-stage method (the Benjamini–Krieger–Yekutieli method) in order to adjust the p-values obtained from the Wilcoxon signed-ranks test, with the final aim to control the false discovery rate. The authors experimented six intra-domain and also 24 cross-domain scenarios by starting from the assumption that in all the data there is shared knowledge that can be used to solve problems in different domains, as all the data can be decomposed into a linear combination of sine and cosine waves; as results, the researchers found only two source–target domain pairs in which the transfer learning models performed with a negative transfer, that is worse than the reference models; thus, Otović et al. concluded that transfer learning is highly likely to either improve or have no detrimental effect on the predictive performance and on training convergence rate of the models. Furthermore, Otović et al. argued that a model pre-trained on an unrelated task can be better than a randomly initialised model, and that pre-training models proved beneficial and should be considered, even when the two domains appear to be unrelated. These results highlight the potential value of leveraging pre-training techniques, regardless of the apparent dissimilarity between the tasks.

In 5-6 [49] Yang et al. have proposed a methodology that combines technical analysis and sentiment analysis to create predictor variables. These variables are then used - within the framework of the transfer learning - with a Least Absolute Shrinkage and Selection Operator-LSTM (LASSO-LSTM) model to forecast the direction of stock prices. By leveraging the advantages of the LASSO algorithm, the authors argued that LASSO-LSTM method can effectively process stock data sets, because it is able to overcome the issue of the high correlated relationship among input variables, as it introduces penalized function that is aimed at compressing the irrelevant variables to 0 (thus by eliminating multicollinearity). To gather data, the researchers scraped financial content from the investing.com website; more in depth, the authors employed transfer learning using the Finbert method (financial sentiment classification problem with Bidirectional Encoder Representations from Transformers) to convert textual data in 4 sentiment indicators, and using the TTR package to calculate 37 technical indicators based on historical daily price data. With the aim to evaluate the model robustness and to mitigate the impact of random factors, after repeating the experiment 30 times - for six different models - the authors calculated not only the accuracy, but also the standard deviation of accuracy as a high standard deviation value indicates a lack of stability and poor robustness in the model. Moreover, Yang et al. used the Wilcoxon signed rank test to determine if there is a significant difference between the prediction results of two compared models. The experiment findings of this study demonstrated that the LASSO-LSTM model, incorporating technical and sentiment indicators, achieves an average accuracy improvement of 8.53% compared to the standard LSTM model. Moreover, this study confirmed - as stated in previous studies - that utilizing historical transactions and financial sentiment data can capture crucial information that influences stock movement. Additionally, the study points out that effective variable selection enables the retention of key variables and improves the predictive performance of the model. As challenges and future directions, Yang et al. stated that employing techniques to fine-tune the hyperparameters of the model, such as the genetic algorithm or particle swarm optimization, could be beneficial to achieve superior results.

In 5-7 [50] He et al. have pointed out how much is impractical the assumption on the access to extensive training data when it comes to predicting stock movement; indeed, in stock markets, numerous stocks lack sufficient historical data, particularly for companies that recently went public through an initial public offering. To tackle the challenge of limited availability of data, the researchers have developed deep transfer learning models on 11 features derived from open, high, low and close daily stock prices. The authors have also stated that while transfer learning demonstrated effectiveness in computer vision and natural language processing domains, its application in stock movement prediction remains limited, accordingly to a cited previous study [51] in which, with dissimilarities between the source and target datasets, a detrimental impact on the performance of the deep learning model has been highlighted. The proposed models have been compared against state-of-the-art prediction models and the experimental results, conducted on three publicly available datasets, demonstrated that the proposed methods significantly enhance the performance of deep learning models when confronted with limited training data scenarios. As challenges and future directions, the researchers stated the intention to develop adversarial samples, generating by small perturbations, to be used in an adversarial evolution of their architecture, to create an attention mechanism that will facilitate the transfer of information from the source dataset to the target dataset, and to investigate how various attention mechanisms impact the model's performance.

In 5-8 [52] Borrageiro et al. have presented use of online transfer learning in the context of a trading agent for digital assets. The trading agent utilized an echo state network, a robust feature space representation, to extract valuable information. The output of this network has then fed into a recurrent reinforcement learning agent, in order to let this agent able to learn how to effectively trade the XBTUSD (Bitcoin versus US Dollars) perpetual swap derivatives contract on an intraday basis (5 minutes). Over a period of approximately five years, the agent achieved a total return of 350% after accounting for transaction costs. This results highlights the agent's ability to effectively capitalize on funding opportunities and generate substantial returns for the given timeframe.

In 5-9 [53] Borrageiro et al. have investigated the potential of online inductive transfer learning by transferring feature representation from a radial basis function network, consisting of Gaussian mixture model hidden processing units, to a direct, recurrent reinforcement learning agent. The main focus is on applying this agent in a trading experiment, on a daily basis, involving major currency pairs in the spot market. The authors have taken into account transaction costs and funding costs to accurately evaluate the agent's performance. The agent received information about profit and loss sources, including currency market price trends, through a quadratic utility function, which has enabled it to get specific positions. The agent's performance has achieved an annual compound return of 9.3%, taking into account execution and funding costs, over a 7-year test period. Thus, this agent has proved its capability to generate consistent returns.

In 5-10 [54] the researchers have presented a transfer learning-based model, with adversarial training, in order to tackle the challenge of weak generalization in stock movement prediction. Adversarial training involves incorporating adversarial examples, which are created by an adversary architecture, with the aim to introduce small perturbations - to clean samples – for increasing generalization's capability and robustness of the model; the transfer learning process has been proposed by incorporating an automatic data selection, in order to consider only relevant relational stock data for predicting the movement of target stocks. The experimental results, by using - as input - technical indicators derived by daily data, demonstrated that the proposed framework outperforms various competitive baselines classifiers and achieves state-of-the-art performance. As future direction, the researchers plan to develop a forecasting scheme that not only achieves a high number of accurate predictions, but also selects the most promising stocks that offer the highest expected revenue, according to the idea that making the right prediction in stock movement can result in a small profit, while an incorrect prediction can lead to substantial losses; furthermore, as the researchers highlight that market sentiment and political events can significantly impact stock movements, they plan to incorporate - in their model - external data derived from news, financial indicators, and political events in order to improve prediction capabilities.

In 5-11 [55] Chen et al. have developed transfer learning-based deep reinforcement learning algorithms on daily closing price and opening price stocks as input variables. The authors opted to choose subsets of constituent stocks that have varying relationships with the Chinese index stocks, as potential candidates for training. To accomplish this, Chen et al. employed two measures to assess the relationship: Pearson correlation and consistency between the neural network inputs and outputs. Transfer learning has been applied in order to reduce the extent to which the model relies on a large volume of data, and – as consequence - to overcome the computational complexity and training time constraints issues. The authors highlighted that their methodology is able to improve the stability of the deep learning network since transfer learning helps to stabilize the returns of the portfolios. However, Chen et al. stated that they did not extensively discuss the detailed analysis of selecting input features; on this point, they stated that incorporating prior knowledge and field experience, along with employing Bayesian models, Principal Component Analysis, and other methods mentioned in the literature, could positively impact on the performance of the prediction model. Furthermore, they planned to investigate the potential benefits of incorporating econometric models, or time-frequency analysis techniques, to enhance the prediction capabilities and generalization ability of their deep learning models; additionally, according to the idea that when it comes to real trading scenarios, investment strategies encompass various crucial elements such as asset allocation, leverage level, hedge ratio, and outputs beyond transaction behaviour and transaction shares, the researchers planned to develop investment strategies that better align with real-world investment practices and that consider a broader range of factors - for more robust and practical predictions - that incorporate more realistic features of fund investment.

In 5-12 [56] Hirchoua et al. have used an evolutional learning method - by combining general-purpose techniques for neural networks, deep reinforcement learning, transfer learning, and multi-agent learning - to address the challenges and complexity of stock markets on a daily base. The system focused on the size of transferred learning, by utilizing the neural network output to determine the mutation size for each generation, and with the aim to balance the merged knowledge size through evolution. This approach helps maintain a balanced and optimal level of knowledge transfer as the evolution progresses. The evolutionary algorithm played a crucial role in this process because it adapted the internal strategies of the agents and drove the system to train new creative behaviours for the subsequent generations. The resulting agent exhibited promising performances. As future direction, the researchers planned to extend the proposed model by replacing Deep Q-Learning with proximal policy optimization and trust region policy optimization algorithms, in order to enhance the performance and efficiency of the model.

In 5-13 [57] Xu et al. have developed a deep transfer learning approach, using a long-short term memory network, with the aim to examine the effectiveness of transfer learning in predicting stock prices. The study selected the Shanghai Securities Composite Index as the source data, and focused on the A-share "ICBC" as target domain data. The experiment – based on closing price data of the past 8 days as input data for training the model – was aimed at predicting the stock price of the "Industrial and Commercial Bank." The researchers conducted experiments to analyze the factors that influence the model's learning performance. One experiment involved fine-tuning the entire network without freezing any layers, while another one frozed the first two LSTM layers and only trained the newly added fully connected layer. The results indicated that, in transfer learning, retraining the entire network often yields better results than training only the fully connected layer. This study also showed that using the composite index - as the source domain - yields better results compared to most individual stocks as the source domain. This finding may be attributed to the fact that the Composite Index reflects the overall changes in individual stock prices. Thus, the experimental results demonstrated the high practical value of transfer learning in stock price prediction, as transfer learning proves to be effective in addressing the challenge of poor prediction accuracy caused by small datasets. However, for datasets with a relatively large amount of data, the study highlighted that accuracy improvement is not significant. As future direction, the researchers planned to go deeper into the factors influencing the selection of source domain data.

In 5-14 [58] Lv et al. have developed a hybrid transfer learning model, by combining transfer learning, variational mode decomposition and bidirectional long-short term memory network, to forecast the intraday (30 minutes) closing price movements of Shanghai Securities Composite Index. The variational mode decomposition has been used in order to decompose the time series into multiple harmonic sub-sequences, with the aim to reduce the complexity of signal frequency domain. By comparing various models, the authors provide evidence that their proposed model exhibits outstanding performance, as it surpasses other compared models in terms of predictive error.

In 5-15 [59] Dong and Liu have worked on sentiment classification from financial texts. They recognized the importance of determining whether a financial text carries a positive sentiment, as it can provide valuable insights into current investment sentiment. Nevertheless, the authors highlighted that sentiment classification of financial text is characterized by the scarcity of annotated data, as a notable challenge distinct from sentiment classification in other domains; thus, Dong and Liu stated that, as a consequence, the application of sentiment classification to financial texts has been considerably limited. To address this issue and to enable the utilization of deep learning techniques for sentiment analysis in financial texts, the researchers developed an approach aimed at employing cross-domain learning methods to facilitate the transfer of sentiment classification knowledge from other domains to the domain of financial texts. The proposed sentiment classification transfer learning method – evaluated by using an open-source dataset (Amazon books, DVDs, electronics, and kitchen appliances) as the source domain for cross-domain learning - yielded improvements in classification accuracy when compared to non-transfer learning approaches. As future directions, the researchers planned to investigate other NN architectures for the sentiment classification task.

In 5-16 [60] Fons et al. have showed the effectiveness of using a transfer learning-based LSTM model for stock classification on daily data. The approach involved pre-training a model on the complete set of stocks, from the S&P500 index, to extract universal features. Additionally, the paper introduced the data augmentation technique, by applying random transformations on the input data and by augmenting the feature vector. A comparison has been made between augmentation in the feature space and the traditional approach of augmenting the time-series in the input space. The results reveal that augmentation methods in the feature space lead to a 20% increase in risk-adjusted returns, compared to models trained with transfer learning but without augmentation. Thus, this study highlights that incorporating transfer learning into a stock classification task improves financial performance compared to training a neural network from scratch. Furthermore, the research indicates that using a training loss that combines a classification objective, with the maximization of returns, enhances risk-adjusted returns when compared to using a single cross-entropy loss. As future direction, Fons et al. planned to explore the learned representations of the pre-trained model, in order to investigate the explainability of these representations, as interpreting the underlying patterns, and characteristics captured by the model's learned representations, allows for a deeper understanding of how the model processes and represents the input data.

In 5-17 [61] Che et al. have developed a transfer learning-based LSTM model for daily stock prediction (uptrend or downtrend) based on sentiment classification. The focus was on utilizing financial news to gain better insights and achieve more accurate predictions of stock prices direction, with a specific emphasis on one-day ahead forecasting classification. Through the utilization of pre-trained word embedding techniques, and a classification layer, the proposed method demonstrated robust success in predicting stock prices. The experimental results indicated that the approach outperforms previous methods, in terms of accuracy, and exhibits advantages in adapting to different datasets.

In 5-18 [62] He et al. have developed two multi-source transfer learning methods, called Weighted Average Ensemble for Transfer Learning (WAETL) and Tree-structured Parzen Estimator Ensemble Selection (TPEES). The proposed approaches, evaluated on financial time series extracted from stock markets, were aimed at addressing the challenge of insufficient training data when predicting stock prices in financial markets. The input vector is constituted by a 22-day historical close price of a stock, while the output vector is the 1-day ahead stock price. WAETL utilizes an average ensemble approach and aggregates the output of the target model by averaging the outputs of each model from the model pool; by means of this simple averaging technique the authors tried to prevents overfitting and to generate a smoother ensemble model. However, not all models - in the model pool - have an equal influence on the target model; thus, to address this issue, WAETL method assigned weights to the source datasets based on their similarity to the target dataset. Various distance functions, including Correlation alignment loss, wasserstein distance, dynamic time warping, and pearson correlation coefficient, have been used to calculate the similarity between the source and target domains, and the similarity values have been transformed into weights using a specific function. The second method have involved pre-training deep learning models that have been fine-tuned with the target dataset, and stored in the model pool. The Tree-structured Parzen Estimator (TPE) algorithm has been employed to select the best models to be used in the ensemble; thus, the average of the ensemble model was calculated with respect to the outputs of the selected models, from the model pool. The experimental results demonstrate that TPEE outperforms other baseline methods in the majority of multi-source transfer tasks. As future direction, the researchers intend to expand their models to incorporate negative correlation, and technical indicators derived from stock market data, with the aim to capture and utilize a broader range of information, allowing for more accurate and insightful predictions.

In 5-19 [63] Koshiyama et al. have reviewed Artificial Intelligence, Machine Learning, and related algorithms in future Capital Markets. Among several analysed forms of learning, the authors specifically focused on two learning paradigms: transfer learning and meta learning, as these paradigms are aimed at capturing knowledge learned from multiple tasks and are aimed at transferring it to new, unseen tasks, with the final aim to accelerate training, prevent overfitting, and enhance the final performance. In their study, Koshiyama et al. highlighted the conceptual difference between transfer learning and meta learning: within the transfer learning framework the knowledge is transferred from a pre-trained model (or a set of models), to a new model, by encouraging the new model to have similar parameters; on the other hand, meta learning involves abstracting and sharing the learning method (learning rule, initialization, architecture, etc.) across tasks; in other words, in transfer learning a pre-trained model is adapted to a new task, while in meta learning a pre-trained optimizer is transferred across problems. In both cases, the common approach is to train a Deep Neural Network that can be reused later. As challenges and future directions, the researchers pointed out that the influence of algorithms, on future Capital Markets, will be driven by several factors: first of all, by the development of self-programming machine learning algorithms that possess the capability to dynamically adjust to the specific characteristics and conditions of their target markets, as this adaptive nature allows for more efficient and effective decision-making; secondly, by the collaboration between algorithms derived from computational statistics, artificial intelligence, and complex systems, as this interdisciplinary approach brings together diverse methodologies and expertise to tackle complex market dynamics; thirdly, by the increasing utilization of Big Data and alternative datasets, as these novel data sources offer unique insights and potential competitive advantages for enhancing the accuracy of market analysis; fourthly, by the adoption of learning techniques which are able to address concerns regarding data privacy and security, while still leveraging the collective intelligence of multiple data providers; fifthly, by the issues of interpretability, ethics, and legality, as understanding how algorithms make decisions - and ensuring they adhere to ethical and legal guidelines - are very relevant aspect for the world development. Thus, Koshiyama et al. argued that exploring the above research drivers will contribute to shaping the future impact of algorithms in Capital Markets and pave the way for advancements in the field.

In 5-20 [64] Nakagawa et al. have implemented factor-based statistics, as effective indicators for predicting stock returns, by creating a multi-factor investment strategy that remains consistent, over a long period, using a transfer learning-based supervised machine learning model. The authors highlighted that, although machine learning methods are increasingly used in stock return prediction, accurately inferring stock returns is difficult, and using complex machine learning techniques without caution can lead to overfitting - on current data - and poor performance on future data. To address these issues, Nakagawa et al. proposed a framework for stock return prediction called Ranked Information Coefficient Neural Network (RIC-NN), with the aim to mitigates overfitting, by incorporating three ideas: a nonlinear multi-factor approach, stopping criteria based on ranked information coefficient, and deep transfer learning across multiple regions. The researchers evaluated RIC-NN using a medium-term investment cycle, where investments are made on a monthly basis by using 20 variables that encompass fundamental and statistical factors. Furthermore, they explored information aggregation across different markets within the MSCI indices, specifically focusing on transfer learning between the North America (NA) region and the Asia Pacific (AP) region. Comparative experiments demonstrated that RIC-NN outperforms both off-the-shelf machine learning methods, and the average return of major equity investment funds, over the past fourteen years. The experimental results, focused on transfer learning, reveal that the NA data can be effectively utilized to predict future returns in the AP market, but the reverse is not true. This supports the hypothesis of an asymmetric causal structure between the two markets. As future research directions, Nakagawa et al. pointed out the relevance to develop more sophisticated portfolio strategies, according to the importance of explicitly considering risk in portfolio selection, as stated in portfolio theory; in order to address this direction, the researchers planned to explore the combination of their method with methodologies such as Subset Resampling Portfolio or Ensemble Growth Optimal Portfolio. Furthermore, Nakagawa et al. planned to build new models, such as LSTM with the attention mechanisms, in order to capture the time evolution of the stock market and the long-range interactions.

In 5-21 [65] Fang et al. have proposed a transfer learning-based approach where several smaller neural sub-networks were combined into a large network, with the aim to fine-tune the prior knowledge captured by the smaller sub-networks. As finance exhibits a high level of instability because of the presence of non-stationary noise, the authors highlighted that this is a significant challenge for powerful methods employed in pattern recognition and natural language processing, as the above methods, which typically yield impressive results in other domains, tend to perform poorly when applied to the field of finance; indeed, the researchers pointed out that DNNs and LSTM tend to overfit non-stationary noise, and that the above models - also implemented with generalization techniques - performs well only when the noise in the data is stationary; but, when the noise is non-stationary, the exceptional fitting capability can become a drawback, as the training set may contain a trap that misleads the models, while the true knowledge in the test sets lies in the opposite direction; in other words, the models fails to perform well when tested with a different type of noise in the testing set with respect to the training set. As in financial time series the noise is predominantly non-stationary, this exacerbates the challenges faced by the above models. As a result, the researchers stated that DNNs and LSTM becomes nearly useless in the context of current quantitative trading algorithms, as this field - closely intertwined with human sentiment and intelligence - is known for its inherent volatility and instability. Thus, the researchers developed a model called the prior knowledge network (PK), constituted by several sub-networks, merged into a large network with a fine-tuning purpose on the prior knowledge captured by the smaller sub-networks. The authors utilized two real datasets (the CSI500 Index and CSI300 Index, which are widely traded indices in the A-share market) and used the previous 50 minutes' index return series for binary classification of the subsequent time period, conducting tests for various durations ranging from 1 minute to 12 minutes. Fang et al. repeated the experiments 10 times and reported the mean performance, along with the standard deviation, to mitigate the impact of randomness. The results demonstrated that the proposed model outperformed traditional methods in terms of speed and accuracy when applied to real financial datasets.

In 5-22 [66] Miao et al. have developed a bidirectional LSTM model, transfer learning-based, to perform sentiment classification tasks in the financial domain. The paper also presented a dynamic financial knowledge graph visualization system to capture data changes and trends over time. The model has been trained on data through web crawling, which encompassed various information related to A-share listed companies. The above data included basic information, and profiles, of A-share listed companies, as well as news, articles, announcements, and research reports related to these companies. The results indicate that transfer learning is effective, particularly when the data is limited or scarce. As challenges and future direction, the researchers planned to get better the dynamic knowledge graph, by incorporating time-series relationships as auxiliary features for relationship prediction. Moreover, Miao et al. pointed out that the models are currently trained sequentially, leading to a strong dependency of the latter models on the outputs of the previous models, and that this dependency can result in error propagation and can negatively impact the performance of downstream models. Thus, to address this issue, the researchers also planned to combine multiple tasks and training them together. This approach should be able to enhance the robustness of the model and to mitigate the potential negative effects of error propagation.

In 5-23 [67] Dhar et al. have developed a transfer learning-based computer vision representation of multiple financial time series, by using images. This approach draws inspiration from deep learning techniques used in computer vision. The authors followed two main research objectives; firstly, Dhar et al. converted time series, which is commonly used in Finance and other fields, into a problem that can be tackled using deep learning methods such as convolutional neural networks. This transformation allowed to leverage the power of deep learning in addressing the modified problem; secondly, the researchers explored the concept of transfer learning, wherein the learning process takes place on synthetic simulated data that reflects a finite set of relationships among the time series. The knowledge acquired from this simulated data has been then directly applied to the real-world application domain of finance. As result, Dhar et al. argued that there must exist a discernible relationship between the simulated and real-world data for transfer learning to be effective.

In 5-24 [68] Nguyen et al. have highlighted that, in recent years, machine learning techniques, particularly deep learning methods like recurrent neural networks (RNNs) with LSTM cells, have gained popularity for their ability to handle sequential data in different domains, such as in financial time series prediction tasks; thus, in their study, those researchers employed transfer learning-based LSTM cells, on daily closing prices, for sequence learning in financial market predictions, as they stated that the transfer learning methodology is able to address the overfitting problem - caused by a limited number of training samples - by leveraging a large source of data obtained from various stocks. Thus, the authors trained a base LSTM, by mixing several stocks data, and developed a fine-tuning step using a small amount of target data. The results demonstrated that the proposed approach is able to outperform other baselines approaches in terms of average prediction accuracy. In order to define next challenges and possible future research directions, Nguyen et al. stated that their model has a large number of parameters, making the training process time-consuming and computationally demanding; moreover, the authors highlighted that selecting an appropriate number of stocks, for the source data of the base LSTM, also influences predictive accuracy. Additionally, they stated that the model's profitability could further be enhanced by means of an optimized trading system that integrates multiple sources of data (including numerical data, like returns), and market sentiment information.

In 5-25 [71] Merello et al. have formulated the stock price prediction problem as a transfer learning–based regression task, specifically predicting market returns; indeed, the researchers argued that stock prediction classification approach typically provide signals for individual stocks independently, without considering the performance of other assets; as a result, the predictions cannot be used to correctly select the best performing stock - among the other ones - for investing purposes; indeed, even if two assets are correctly predicted as "buy," their fluctuation strengths may differ significantly, and investing the entire capital on the most performing asset would yield higher returns. The regression-based approach estimated the direction and also the magnitude of price changes for each stock, enabling identification of the most performing assets for portfolio building purposes. Thus, Merello et al., by comparing the results of their regression approach with respect to a binary classification one (on trends categorized as 'up' or 'down'), demonstrated the effectiveness of applying transfer learning techniques to stock market prediction. To achieve this, they considered the information available at a given time, including technical indicators and news articles, as factors influencing the trends. Findings demonstrated that the returns predicted by the regression approach offer more meaningful insights compared to the 'buy' or 'sell' signals provided by the classification approach, according to the idea that regression results can be leveraged to enhance financial profits through an investment strategy focused on the most performing assets. As future research directions, the researchers highlighted the need to benchmark the benefits of their regression approach against state-of-the-art methods, by comparing their model with existing methods in order to measure the improvements and advantages offered by their proposed approach.

In 5-26 [72] Man et al. have highlighted the influence of market sentiment on various aspects of financial markets, such as price trends, trading volumes, volatility, and risks; as a result, they emphasized that several trading strategies, developed on the base of the findings of financial sentiment analysis (FSA), leading to significant returns; the researchers also reported example of the above topic relevance, by highlighting that major news vendors, such as Thomson Reuters, have incorporated sentiment analysis into their services (Thomson Reuters News Analytics – TRNA - scores) that assess the polarity of news content; moreover, the researchers pointed out the importance of Natural Language Processing techniques on the topic, because these techniques facilitates the analysis of the financial text data. As possible future directions for FSA research, the paper pointed out the relevance of information combination, from different data sources, to enhance analysis, and the importance of transfer learning technique as path for more effective sentiment analysis.

In 5-27 [73] Jeong and Kim have pointed out that stock trading presents several challenges because financial data is often limited, volatile, and contains uncertain patterns, making decision-making difficult. To address this, the researchers proposed a transfer learning-based reinforcement learning model aimed at addressing the problem of insufficient financial data, by selecting component stocks based on their relationship with an index stock; thus, Jeong and Kim highlighted that the above approach prevents overfitting and provides useful information, leading to a better understanding of the financial data. However, researchers stated that their study has certain limitations because it analyses data on closing trade prices, so it overlooks other important market factors like volatility, and because it focuses on a single stock, but it would be valuable to extend it to a portfolio of stocks.

In 5-28 [74] Boytcheva and Tagarev have argued that finding investment opportunities can be overwhelming, especially when considering the vast number of companies not publicly known; indeed, these lesser-known companies may hold the potential for technological innovation, market disruption, and excellent investment prospects; but, manually selecting through millions of options is impractical. To address this challenge, the researchers proposed a recommendation model for alternative company investments, by means of a clustering approach based on the transfer learning idea. To validate the approach, the researchers conducted experiments on a dataset comprising 7.5 million companies, with a specific focus on startups and investments made within the past three years. Findings show that the approach provides investors with effective guidance in discovering investment options, while minimizing the need to manually explore an extensive pool of companies. As future direction, the authors of the study highlighted that the next step could be the development of a more robust evaluation methodology; furthermore, in terms of feature engineering, Boytcheva and Tagarev stated that a beneficial direction could be to address the company descriptions using Natural Language Processing techniques; this could involve employing a neural network-based approach capable of performing meaningful text processing, with the objective to identify semantic similarities, among company descriptions, and with the aim to represent them numerically, enabling their integration into current algorithm; this NLP-based processing would enhance the quality and relevance of the input data, contributing to improved outcomes.

In 5-29 [75] Tripathi have pointed out that the finance industry faces challenges regarding data availability and individualizing services beyond existing customers; moreover, the researcher highlighted that - compared to consumer applications - the finance and banking sector has limited data to personalize their offerings, and that one prominent issue is data restriction, where financial institutions have limited access to data due to privacy concerns and regulatory constraints; thus, Tripathi argued that the above limitation hinders the development of sophisticated ML models that rely on large amounts of data, and emphasized that transfer learning emerges as a promising technique, for overcoming this challenge, because this approach allows financial institutions to utilize data from similar domains to improve their ML models. As future research direction, Tripathi pointed out that AI and machine learning interpretability are rapidly evolving fields of research, experiencing significant changes and expansions at present, and that achieving interpretability in machine learning models remains a considerable challenge, particularly in the finance and banking industry. As consequence, the researcher argued the need to go deeper on the above topic because for these industries it is crucial to prioritize the development and adoption of explainable models that provide a clear understanding of the decision-making process before implementing machine learning solutions in real-world applications.

In 5-30 [76] Nayak et al. have argued that the extraction of key information, such as important phrases and numerical data, plays a crucial role in banking and financial processes, but they also pointed out that openly available datasets - in the financial sector - are limited in terms of labelled data; as consequence, the researchers stated that in the banking and financial industry there is the need to address the challenges associated with developing a data extraction system for low-resource datasets. To tackle this issue, Nayak et al. experimented a Bi-directional Long Short-Term Memory model, and observed that, by applying transfer learning from an out-of-domain dataset, the accuracy of several data extraction tasks results significantly improved. Thus, the researchers argued that combining transfer learning with domain-specific knowledge enhances entity recognition in low-resource settings. As future directions, the researchers stated the interest to improve their transfer learning-based model by integrating their world embedding process with other embedding ones, such as ELMo Embeddings [78] or Bert Embedding [46].

### 5.2.1 Answers to the research question #1

Transfer Learning has been applied for financial market predictions in different ways:

- several researchers have developed transfer learning models in order to overcome the financial data scarcity issue [21, 43, 48, 50, 57, 59, 62, 66, 73, 75, 76]; as example, by starting from the observation that the common financial markets characteristics are able to explain the capability of technical indicators for providing working financial forecast signals, some scholars used transfer learning models in order to extract the general pattern underlying financial data of different securities markets; a recent study [24], by highlighting that <<current approaches in trading strategies treat each market or asset in isolation, with few use cases of transfer learning in the financial literature>> showed that a transfer learning model, which can learn a trading rule directly from a large-scale stock dataset and which is able to fine-tune it on a dataset that has the trading rule included, <<improves financial performance when compared to training a neural network from scratch>> [24]; moreover, another recent study stated that <<transfer learning based on 2 data sets is superior than other base-line methods>> [23].

- Literature have highlighted the role of transfer learning as a method for accelerating the training [64, 65], and for discovering asymmetric causal structure between different domains; as example on the last point, Nakagawa et al. [64] showed that the North America stock price data can be effectively utilized to predict future returns in the Asia-Pacific market, but the reverse is not true. This supports the hypothesis of an asymmetric causal structure between the above two domains.

- Several studies [22, 55, 59, 63, 68, 80] have also pointed out the transfer learning capability to address the overfitting problem, by means of leveraging source of data belonging to different domains; as example, Otović et al. [22] suggested that utilizing a pre-trained model, from an unrelated task, can yield better results compared to initializing a model randomly; in other words, the research findings emphasize the benefits of pre-training models, and advocate for their consideration even when the tasks seem unrelated.

- There are contrasting literature results on the effectiveness of the transfer learning idea to exploit the prediction power of some pre-trained CNNs architectures,: indeed, a study [25] stated that <<a 2D-convolutional neural network (2DCNN) outperformed the other popular networks (VGG16, RestNet50, and random forest) for stock price movement prediction using candlestick chart data>>; on the contrary, a recent work [32] highlighted the effectiveness of a pre-trained CNN video prediction technique (<<mirrored VGG16-LSTM>>) aimed at first visualizing the multivariate time-series data like a sequence of images (thus forming a video), and then at predicting future image frames.

### 5.2.2 Answer to the research question #2

Analysed Literature highlighted the following challenges and potential future directions of the transfer learning methodologies for financial market predictions:

- in order to improve the transfer learning process, literature have highlighted the relevance to go deeper into the factors influencing the selection of source domain data [57, 80], by considering also the number of domains to be used [68];

- literature suggested to address the possible error propagation issue related to the usual transfer learning process, as it involves sequential training of models that can adversely affecting the performance of downstream models [66].
- literature have suggested to investigate the impact of different learning mechanism, such as the attention mechanisms as example, on the transfer learning performance [50].

# 6 Conclusions

Literature highlighted that financial time series data pose significant challenges for accurate stock price prediction, because these data are characterized by noise and susceptibility to news; traditional statistical methodologies made assumptions, such as linearity and normality, which are not suitable for the non-linear nature of financial time series; on the other hand, machine learning methodologies - and in particular the neural network ones - are able to capture non linear relationship in the data. Transfer Learning, as a method aimed at extracting knowledge from source tasks and applies the knowledge to target tasks, let to leverage the knowledge - of previously trained models - in order to achieve better performances than traditional ML models.

As existing reviews are mainly focused on neural network architectures - for financial prediction - with very little emphasis on the transfer learning methodology, this study analyzes the current state of the art on the transfer learning methodology, for financial data predictions, by developing a systematic review paper that addresses the relationship between transfer learning and financial forecasting. This analysis shows that several researchers have developed transfer learning models in order to overcome the financial data scarcity issue, and as a method for accelerating the training, and also as a methodology for discovering asymmetric causal structure between different domains; furthermore, this study points out the transfer learning capability to address the overfitting problem, by means of leveraging source of data belonging to different domains. As challenges and potential future directions of the transfer learning methodologies for financial market predictions, this analysis highlights the relevance to go deeper into factors influencing the selection of source domain data, by also considering the number of domains to be used; moreover, this study points out the need both to address the possible error propagation issue related to the usual transfer learning process, and to investigate the impact of different learning mechanism - such as the attention mechanisms as example - on the transfer learning performance.